\documentclass[a4paper,fleqn]{cas-dc}
\usepackage[authoryear]{natbib}
\usepackage{siunitx}

\begin{document}

\let\WriteBookmarks\relax
\def\floatpagepagefraction{1}
\def\textpagefraction{.001}

\shorttitle{Resonant Chain Stability}
\shortauthors{Goldberg, Batygin \& Morbidelli}

\title[mode = title]{A Criterion for the Stability of Planets in Chains of Resonances}
\date{\today}

\author[1]{Max Goldberg}[orcid=0000-0003-3868-3663]
\cormark[1]
\ead{mg@astro.caltech.edu}

\author[2]{Konstantin Batygin}[orcid=0000-0002-7094-7908]

\author[3]{Alessandro Morbidelli}

\cortext[1]{Corresponding author}

\affiliation[1]{organization={Department of Astronomy, California Institute of Technology}, addressline={1200 E. California Blvd}, city=Pasadena, state=CA, postcode=91125, statesep={}, country=USA}

\affiliation[2]{organization={Division of Geological and Planetary Sciences, California Institute of Technology}, addressline={1200 E. California Blvd}, city=Pasadena, state=CA, postcode=91125, statesep={}, country=USA}

\affiliation[3]{organization={Laboratoire Lagrange, UMR7293, Université Côte d’Azur, CNRS, Observatoire de la Côte d’Azur}, addressline={Boulevard de l’Observatoire}, postcode=06304, postcodesep={}, city={Nice Cedex 4},  country=France}

\begin{abstract}
Uncovering the formation process that reproduces the distinct properties of compact super-Earth exoplanet systems is a major goal of planet formation theory. The most successful model argues that non-resonant systems begin as resonant chains of planets that later experience a dynamical instability. However, both the boundary of stability in resonant chains and the mechanism of the instability itself are poorly understood. Previous work postulated that a secondary resonance between the fastest libration frequency and a difference in synodic frequencies destabilizes the system. Here, we use that hypothesis to produce a simple and general criterion for resonant chain stability that depends only on planet orbital periods and masses. We show that the criterion accurately predicts the maximum mass of planets in synthetic resonant chains up to six planets. More complicated resonant chains produced in population synthesis simulations are found to be less stable than expected, although our criterion remains useful and superior to machine learning models.
\end{abstract}



\maketitle

\section{Introduction}
Although compact systems of sub-Neptune planets are abundant, a detailed understanding of their formation remains incomplete. According to most theories of planet formation, planets form in gaseous protoplanetary disks where interactions between the planets and gas are inevitable. These interactions cause inward migration of the planets towards the disk's inner edge and capture them into chains of mean-motion resonances \citep{Terquem2007,Cresswell2008,Ida2008,Ida2010,Cossou2014,Hands2014}. Indeed, we expect that resonant chain systems such as TRAPPIST-1 Kepler-80, Kepler-223, and GJ 876 formed in this way \citep{Mills2016,Luger2017}. 

Yet, population studies of exoplanet systems have revealed that resonant chains are in fact rare and that systems of multiple sub-Neptune planets are typically \textit{not} in resonance \citep{Fabrycky2014}. Thus, either some process prevents the formation of resonances in the first instance, or primordial resonant chains are disrupted after the gaseous nebula dissipates. Recent work \citep{Izidoro2017,Izidoro2021,Goldberg2022} argues for the latter scenario, hypothesizing that widespread dynamical instabilities break the resonances and then a phase of giant impacts sculpts the system. Detailed simulations of such a process produce results matching the observed period ratio distribution, transit multiplicities, and peas-in-a-pod patterns of intrasystem uniformity. However, the mechanism of the instability itself is not well understood from a fundamental level, nor is there a practical way to predict which resonant chain systems are unstable.

The stability of planetary systems has been a topic of research for centuries since the development of celestial mechanics \citep{Laplace1799,LeVerrier1840,Poincare1899}. With the introduction of numerical integration, the Solar System was recognized to be chaotic \citep{Wisdom1983,Roy1988,Laskar1989} and hence unpredictable on gigayear timescales, at least on a quantitative level \citep{Batygin2008,Laskar2008}. Now, the rapidly growing population of exoplanetary systems, and their exotic architectures, has spurred a renewed interest in fully understanding the stability of general systems of planets \citep{Deck2012,Batygin2015b}.

Previous studies, while extensive, have generally focused on two-planet systems \citep{Gladman1993,Deck2013,Hadden2018,Petit2018} and the non-resonant $3+$ planet regime \citep{Chambers1996,Quillen2011,Petit2020,Tamayo2021,Rath2021}. On the other hand, the stability of planetary systems in chains of resonances has received only limited attention. Early work was primarily empirical: \cite{Matsumoto2012} performed numerical integrations of equal mass planets locked into $k$:$k-1$ resonance and found that the maximum number of planets that could be captured into the chain decreased with increasing $k$ and planet mass. Later work \citep{Matsumoto2020} confirmed these conclusions and uncovered the unexpected result that a nominally stable resonant chain could be made unstable by a \textit{decrease} in the mass of either the planets or the star.

On the analytical side, \cite{Pichierri2020} considered an equal-mass three-planet system as the simplest instance of a first-order resonant chain. Through involved perturbation theory, they showed that a secondary resonance between the fastest resonant libration frequency and a difference of the synodic frequencies can drive an instability. Rather than continuing their analytical approach, in this work we simplify their results and generalize to unequal masses and an arbitrary number of planets. Our analysis of the \cite{Pichierri2020} mechanism naturally leads to a criterion for the stability of a resonant chain, and a limit on the planet mass---or alternatively, multiplicity---in a resonant chain. We verify these results numerically on a suite of synthetic planetary systems.

\section{Analytical Estimate of Stability}
We define a resonant chain as a system of three or more planets in which each adjacent pair of planets is locked into mean-motion resonance. One can construct a wide variety of oscillation frequencies from the orbital elements, but important frequencies can be broadly separated into three categories: synodic, resonant, and secular. Synodic frequencies are linear combinations of the mean motions $n_i$ and do not depend on planet masses. Resonant frequencies describe the oscillations of critical resonant angles which, for two-body first-order resonances, take the form
\begin{equation}
    \phi_{i,i+1} = k_i\lambda_{i+1} - (k_i - 1)\lambda_i - \varpi.
\end{equation}
Here, $k_i$ is the resonant index, $\lambda_i$ is the mean longitude, and $\varpi$ is the longitude of pericenter of the $i$-th or $i+1$-st planet. Finally, secular frequencies, which arise from orbit-averaged perturbations, are typically much slower than synodic and resonant frequencies and thus are not considered in this work.

\cite{Pichierri2020} hypothesized that the onset of dynamical instability in compact resonant chain systems is triggered by the commensurability, or near equality, between a resonant libration frequency and a difference of synodic frequencies. Modulation of the resonant angles by synodic perturbations allows the resonant locks to break, leading to chaotic behavior. Our goal is to extend that work to more than three planets of unequal mass.

Consider a resonant chain of $N$ planets with masses $m_1,...,m_N$ and in pairwise first-order resonances of $k_1$:$k_1-1,\ldots,k_{N-1}$:$k_{N-1} - 1$ so that the period ratios are $P_i/P_{i+1} \approx (k_i-1)/k_i$. Studying the secondary resonance of \cite{Pichierri2020} would require writing the Hamiltonian of the entire system. However, we can take a simpler approach by comparing the libration frequencies of the individual resonances to the differences in synodic frequencies throughout the system.

For the purposes of computing libration frequency, we will ignore the contributions of planets that are not in the pair being considered. We verify this assumption below in our n-body simulations. In that case, the angular frequency of libration for the angle $\phi_{i,i+1} = k_i\lambda_{i+1} - (k_i - 1)\lambda_i - \varpi_i$ is approximately \citep{Batygin2015}
\begin{equation}
    \omega_{i,i+1} = \frac{3n_i}{2} \left(\frac{m_1+m_2}{M_*}\right)^{2/3} \left[\frac{((k_i- 1 )^{10}/k_i^4)^{1/9}}{(3(f_\text{res})^2)^{-1/3}}\right]
    \label{eq:omegalib}
\end{equation}
where $M_*$ is the stellar mass and $f_\text{res}\approx -0.8k_i + 0.34$ is a constant derived from Laplace coefficients \citep[e.g.][]{Deck2013}. As discussed in \cite{Pichierri2020}, the $(m_i/M_*)^{2/3}$ scaling is appropriate only at low eccentricities where a shift in the equilibrium point induces a forced eccentricity. At higher $e$, the scaling is $(m_i/M_*)^{1/2}$.\footnote{Specifically, the change in scaling occurs at $e\sim(|f_\text{res}|m/(k^2 M_*))^{1/3}$, where there is a bifurcation in the resonant equilibria in the phase space of the Hamiltonian \citep{Batygin2013}. For the typical systems discussed in this paper, this corresponds roughly to $e\sim 0.03$.}

Synodic frequencies are straightforward to compute. Following \cite{Pichierri2020}, we have
\begin{equation}
    \delta \dot{\lambda}_{i,i+1} = n_i - n_{i+1} = n_i - \frac{k_i - 1}{k_i} n_i = \frac{1}{k_i} n_i
\end{equation}
as the angular frequency of conjunctions of planets $i$ and $i+1$. However, the analysis of \cite{Pichierri2020} specifically identifies the \textit{difference} in synodic frequencies as the slower and more relevant frequency. This is
\begin{equation}
    \Delta\delta\dot{\lambda}_{i,i+1,i+2} = \delta\dot{\lambda}_{i,i+1} - \delta\dot{\lambda}_{i+1,i+2} = \frac{k_{i+1} - k_i + 1}{k_i k_{i+1}} n_i.
    \label{eq:synoddiff}
\end{equation}
Synodic frequencies are typically faster than libration frequencies. Thus, overlap is most likely to occur when the slowest difference of synodic frequencies is commensurate with the fastest libration frequency. We define the characteristic quantity for resonant chain stability to be
\begin{equation}
    \chi \equiv \frac{\min(\omega_\text{syn})}{\max(\omega_\text{lib})},
    \label{eq:chi}
\end{equation}
where the minimum and maximum are taken over all synodic and libration frequencies in the chain, respectively.

So far we have maintained generality, but for simplicity we will now assume that all the resonances have the same index $k$ and the mass of each planet is $m$. Now, the slowest difference of synodic frequencies is $\Delta\delta\dot{\lambda}_{N-2,N-1,N}$ and the fastest libration frequency is $\omega_{1,2}$. Setting $\chi=1$ leads to a maximum mass of planets in the chain of
\begin{equation}
    m_\text{max}/M_* 
    \approx 0.2 \left(\frac{k-1}{k}\right)^{1.5N} k^{1.2} (k-1)^{-6.2} 
    \label{eq:mmax}
\end{equation}
This is an explicit computation of the critical mass identified by \cite{Pichierri2020}.

\section{Numerical Tests}
We ran a suite of numerical experiments to test the validity of Equation \ref{eq:mmax} for different values of $k$ and $N$. We place $N$ planets of mass $m/M_*=3\times10^{-6}$ on orbits $1-2\%$ wide of the $k$:$k-1$ resonance. The semi-major axis of the inner planet is fixed and eccentricity damping is applied to all planets. To ensure sequential capture into resonance, migration is turned on for every planet except the innermost one, using a ratio of migration to eccentricity damping timescales of  $\tau_m/\tau_e = 3\times10^2$. Once the two-body resonant angles begin to librate, we remove migration and eccentricity damping exponentially so that the system settles to its stable multi-resonant state. Typical orbital eccentricities at this point are $\sim 0.01$. Then, we begin exponentially increasing the mass of each planet adiabatically (i.e. with $\tau_m \gg 1/\omega_{1,2}$) in order to increase the libration frequency. Once an instability occurs (defined as any planet orbit becoming hyperbolic) we stop the simulation and record the planet masses. We attempted this process for integer values of $k$ between 2 and 8, and $N$ between 3 and 9. For each pair of $k$ and $N$, we repeated the simulations 10 times with slightly different initial conditions to smooth over the chaotic behavior, although in all cases the scatter was very small. Our simulations use the \texttt{whfast} n-body integrator from the \texttt{rebound} software package and a maximum timestep of $1/20$ of the inner orbital period \citep{Rein2015}. Migration and eccentricity damping are included from the \texttt{reboundx} extension \citep{Tamayo2020a}.

\begin{figure*}[h]
    \centering
    \includegraphics[width=\textwidth]{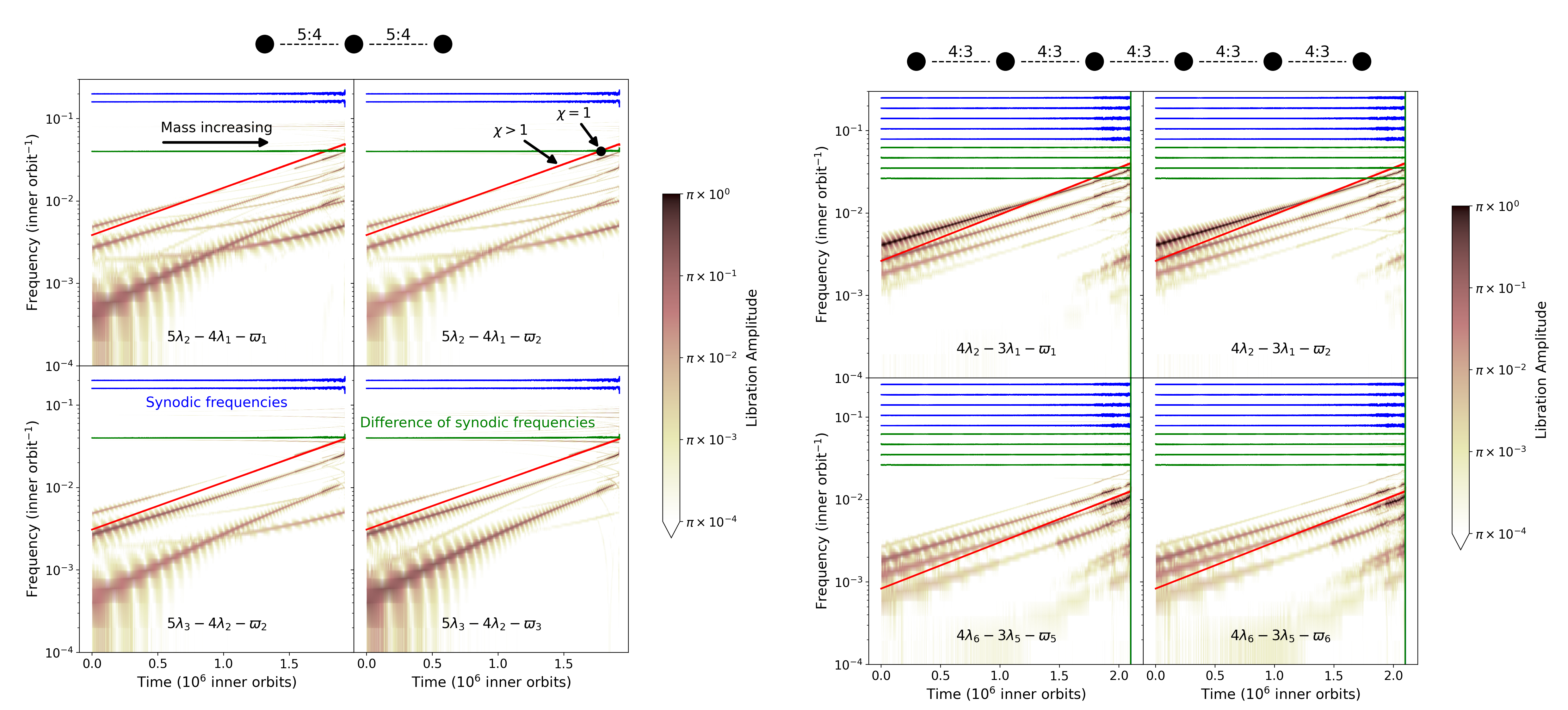}
    \caption{The evolution of resonant libration and synodic frequencies as planet masses are increased until the instability occurs, for two initial planet configurations shown as cartoons above each grid. The left grid corresponds to a system with 3 planets started in 5:4 resonances. Each panel represents one of the four resonant angles; the red colormap is a spectrogram, or the amplitude of the Fourier transform over time, of that angle. Each resonant angle has multiple libration modes, the frequencies of which increase with mass. The bright red lines plot the analytical estimate of the libration frequency from Equation \ref{eq:omegalib}. Horizontal colored lines indicate synodic frequencies: blue lines are the synodic frequencies themselves and the green line is the difference of synodic frequencies (Equation \ref{eq:synoddiff}). The right grid is the same as the left, but with 6 planets in a chain of 4:3 resonance. Only the libration frequencies for the innermost (top) and outermost (bottom) pairs of planets are plotted.}
    \label{fig:3_5}
\end{figure*}

An example of such a simulation is shown in the left panel of Figure \ref{fig:3_5}, which has $N=3$ and $k=5$. As the planetary masses are increased, the libration frequencies increase, but the synodic frequencies remain constant. After $2\times 10^6$ orbits of the inner planet, the resonant angles begin to circulate and the orbital eccentricities grow rapidly until there is a close encounter. Within a few orbits, at least one planet orbit becomes hyperbolic and the simulation ends. The onset of instability happens almost precisely when the highest-frequency mode of the innermost resonant angle (involving $\lambda_1$ and $\lambda_2$), as estimated by Fourier transform of the libration angle, intersects the difference of synodic frequencies. The bottom panels show that the libration of the resonant angle involving $\lambda_2$ and $\lambda_3$ is slower and a resonance with the synodic frequencies does not occur within the simulation timeframe. Figure \ref{fig:3_5} also demonstrates the accuracy of the analytical estimate for libration frequency, which remains within a factor of 2 of the true value throughout the simulation. The libration frequency approximation predicts that the instability will arise at $m_\text{max}/M_*=1.0\times 10^{-4}$, whereas in the simulation the instability comes slightly later, at $m_\text{max}/M_*=1.4\times 10^{-4}$. Nevertheless, the numerically-estimated libration frequencies grow more steeply with mass than the analytical estimate, suggesting that the low eccentricity assumption in Equation \ref{eq:omegalib} has been violated.

A more complicated example is shown in the right panel of Figure \ref{fig:3_5} in which $N=6$ and $k=4$. Here, the frequency structure is more complex and the outermost synodic frequencies are slower. At the first crossing of libration and synodic frequencies, there is a resonant kick and the libration amplitudes increase instantaneously (visible as the blue synodic frequency lines becoming thicker). Upon the equality of the fastest libration frequency and the second-slowest synodic frequency, the resonant angles begin to circulate and the instability is triggered.
Because the instability happens after the libration frequency of the inner planet pair has `overshot' the difference in synodic frequencies of the outer triplet, the analytical maximum mass prediction is an underestimate of the simulation results by a factor of $\sim 2$.

\begin{figure*}
    \centering
    \includegraphics[width=0.9\textwidth]{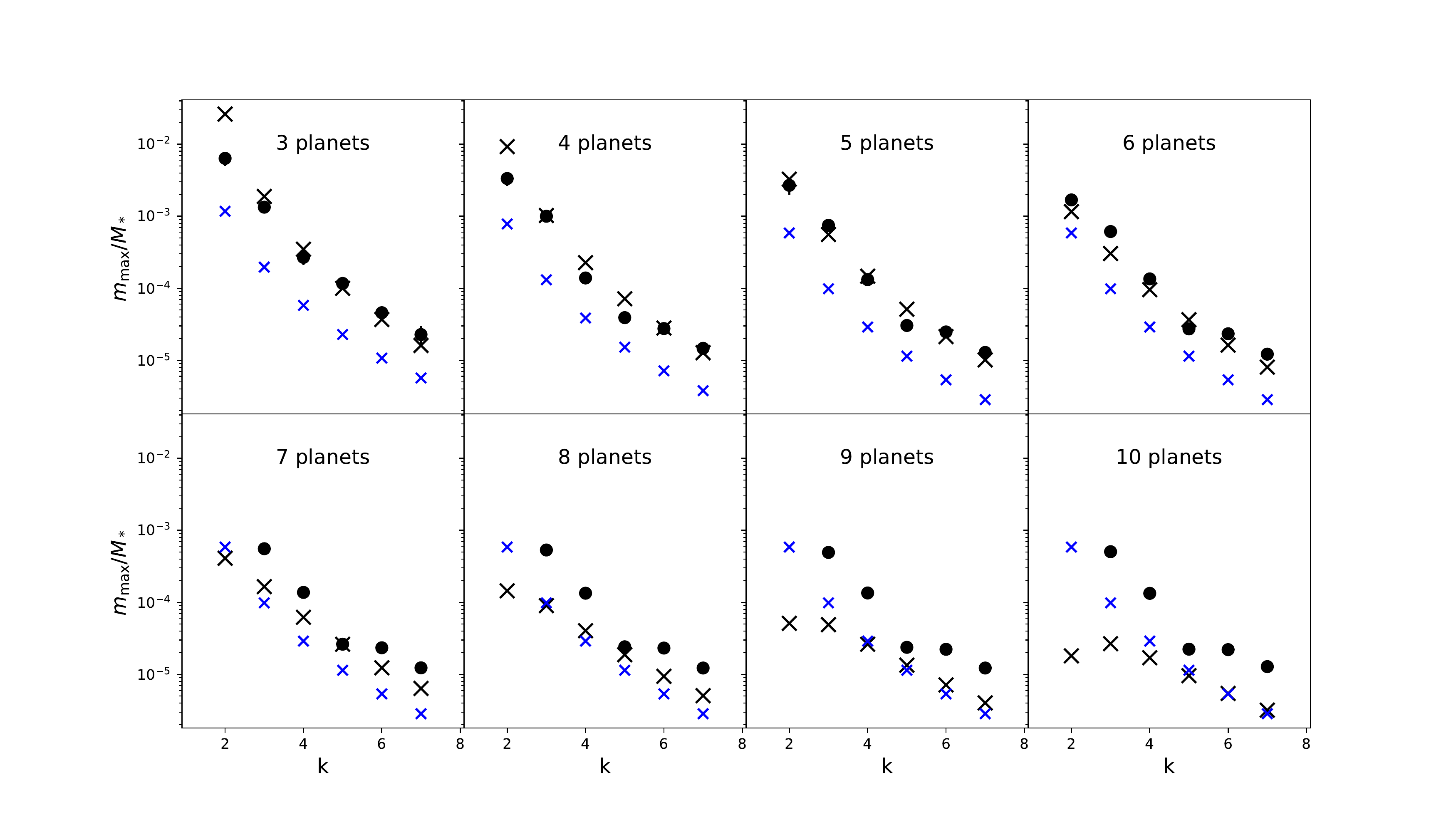}
    \caption{Maximum planet mass in a resonant chain as a function of resonant index $k$, for different planet multiplicities. Black crosses mark the analytical estimate from Equation \ref{eq:mmax}, while dots show the results of our numerical simulations. The smaller blue crosses are the \textit{non}-resonant stability boundary from \cite{Petit2020}.}
    \label{fig:Nk}
\end{figure*}

\begin{figure*}
    \centering
    \includegraphics[width=0.9\textwidth]{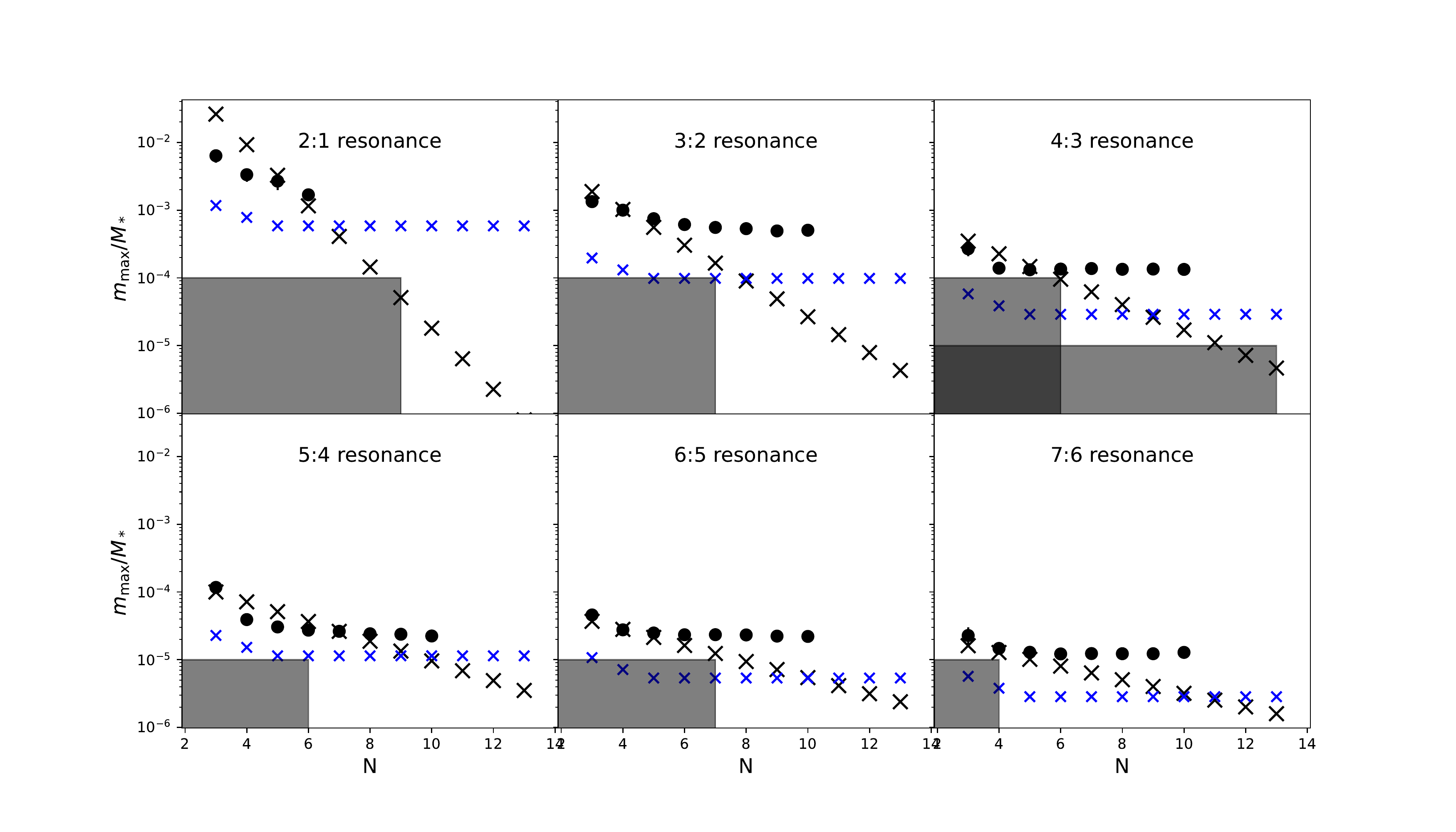}
    \caption{Maximum planet mass in a resonant chain as a function of planet multiplicity $N$, for different resonances. As in Figure \ref{fig:Nk}, black and blue crosses mark the resonant and non-resonant criterion respectively, while dots show the results of our numerical simulations. Gray boxes reflect the implied regions of stability from \cite{Matsumoto2012} and \cite{Matsumoto2020}.}
    \label{fig:kN}
\end{figure*}

The full set of simulations is summarized in Figures \ref{fig:Nk} and \ref{fig:kN}. Figure \ref{fig:Nk} explores how the maximum planet mass varies with resonant index $k$ for constant multiplicity. Our analytical estimate is an excellent fit to the numerical results over a broad range of parameter space. In particular, Equation \ref{eq:mmax} maintains accuracy for values of $k$ between $3$ and $7$, correctly reproducing the downward trend with $k$. This trend is in fact analogous to the Hill spacing stability criterion in non-resonant systems because the semi-major axis ratios are smaller for higher $k$. However, instability in non-resonant systems can be fully accounted for by averaging over synodic terms and considering two-body resonance overlap and three-body resonance diffusion \citep{Petit2020,Rath2021}. Indeed, Figure 2 demonstrates that the non-resonant stability boundary from \cite{Petit2020}, including the $>4$ planets correction, consistently predicts a smaller maximum planet mass than is actually seen in resonant chains. Resonant chains can be stable at separations for which non-resonant systems are unstable because resonant chains reside at a fixed point in the phase space. However, interactions between synodic and resonant frequencies can excite the system away from this fixed point and into the surrounding chaotic region.

Figure \ref{fig:kN} contains the same data but shows how the maximum mass varies with multiplicity $N$ for a constant resonant index. Our analytical estimate predicts an exponential decrease in $m_\text{max}$ with $N$. While this is true for small $N$, the dependence on multiplicity seems to flatten out near $\sim 6$ planets. This may be because the inner resonance is less able to ``communicate'' its frequency to the outer planets for high values of $N$, and as a result, the chain behaves like one with fewer planets. It is worth noting that a similar pattern of saturation, in which stability decreases with $N$ but flattens after $N\gtrsim 5$, occurs in the non-resonant case \citep{Chambers1996}. A somewhat more accurate definition of $\chi$ (Equation \ref{eq:chi}) would therefore consider only adjacent subsystems of $5-6$ planets. However, to maintain simplicity, for this work we will use the previous definition that assumes perfect coupling among all planets.

Figure \ref{fig:kN} also shows poor agreement between our prediction and simulations for the 2:1 resonance. This is likely due to the presence of indirect terms and asymmetric libration in that resonance \citep{Beauge1994}. That is, when eccentricities grow past $\sim 0.03$, the libration centers shift away from $0$ and $\pi$. This is exceeded in our numerical experiments for the 2:1 resonance and our analytical estimates do not consider the asymmetric libration. 

\section{Applications to the Formation of Planetary Systems}
While the results of the previous section are promising, it remains to be demonstrated whether the criterion for resonant chain stability is relevant to the more complex system architectures that are anticipated in the formation of compact super-Earth systems. Here, we apply the criterion to simulations of super-Earth system formation to show that it effectively predicts their long-term stability as well. 

Our test sample is the set of synthetic planetary systems produced in the simulations of \cite{Izidoro2021}. The final systems closely replicate many aspects of the observed sample of compact super-Earth systems. Beyond n-body dynamics, these simulations incorporate orbital migration and eccentricity and inclination damping due to planet-disk interactions as well as pebble accretion. In other words, the simulation suite of \cite{Izidoro2021} constitutes a successful instance of population synthesis. Within the context of these formation simulations, the gas disk dissipates at $t=\SI{5}{\mega yr}$, but the integrations continue until $t=\SI{50}{\mega yr}$ in order to allow for instabilities that were suppressed by the protoplanetary disk to arise. We consider ``stable'' systems to be those that do not experience an instability after $t=\SI{5}{\mega yr}$, and ``unstable'' systems to be those that did experience an instability after $t=\SI{5}{\mega yr}$.

We removed planets with masses below $0.3M_\oplus$ because they tend to interfere with analyzing the chain while not contributing significantly to the dynamics. We also removed systems with a pair of planets that have semi-major axis ratios less than 1.05 because our criterion does not account for the 1:1 resonance. After these cuts, there were 54 unstable systems and 30 stable ones.

The next step is to identify the likely resonances within the chain. We do this by computing the period ratio of adjacent planets. If the ratio is within $3\%$ of a first-order resonance $k$:$k-1$, for $1<k<11$, we assume the planet pair lies in that resonance. If not, we search for second- and third-order resonances with the same method but halve the threshold distance. In the case that no candidate resonance is found, we consider the chain to end at that point. For each planetary system, this process generates a collection of resonant chains separated by secular architecture. Chains with fewer than three planets are discarded because they have no difference of synodic frequencies.\footnote{Specifically, stability for two planets is set by the Hill criterion \citep{Gladman1993,Petit2018}} We then calculate the libration frequency $\omega_{i,i+1}$ for each first-order resonance using Equation \ref{eq:omegalib} (higher-order resonances are ignored) and the difference of synodic frequencies $\Delta\delta\dot{\lambda}_{i,i+1,i+2}$ for each (adjacent or non-adjacent) planet triplet using Equation \ref{eq:synoddiff}. Finally, the stability criterion is computed using Equation $\ref{eq:chi}$. Because some systems contain multiple resonant chains, and an instability in any one of the chains classifies the system as unstable, the system $\chi$ is taken to be the smallest $\chi$ of any of the chains. 

\begin{figure}
    \centering
    \includegraphics[width=1.1\linewidth]{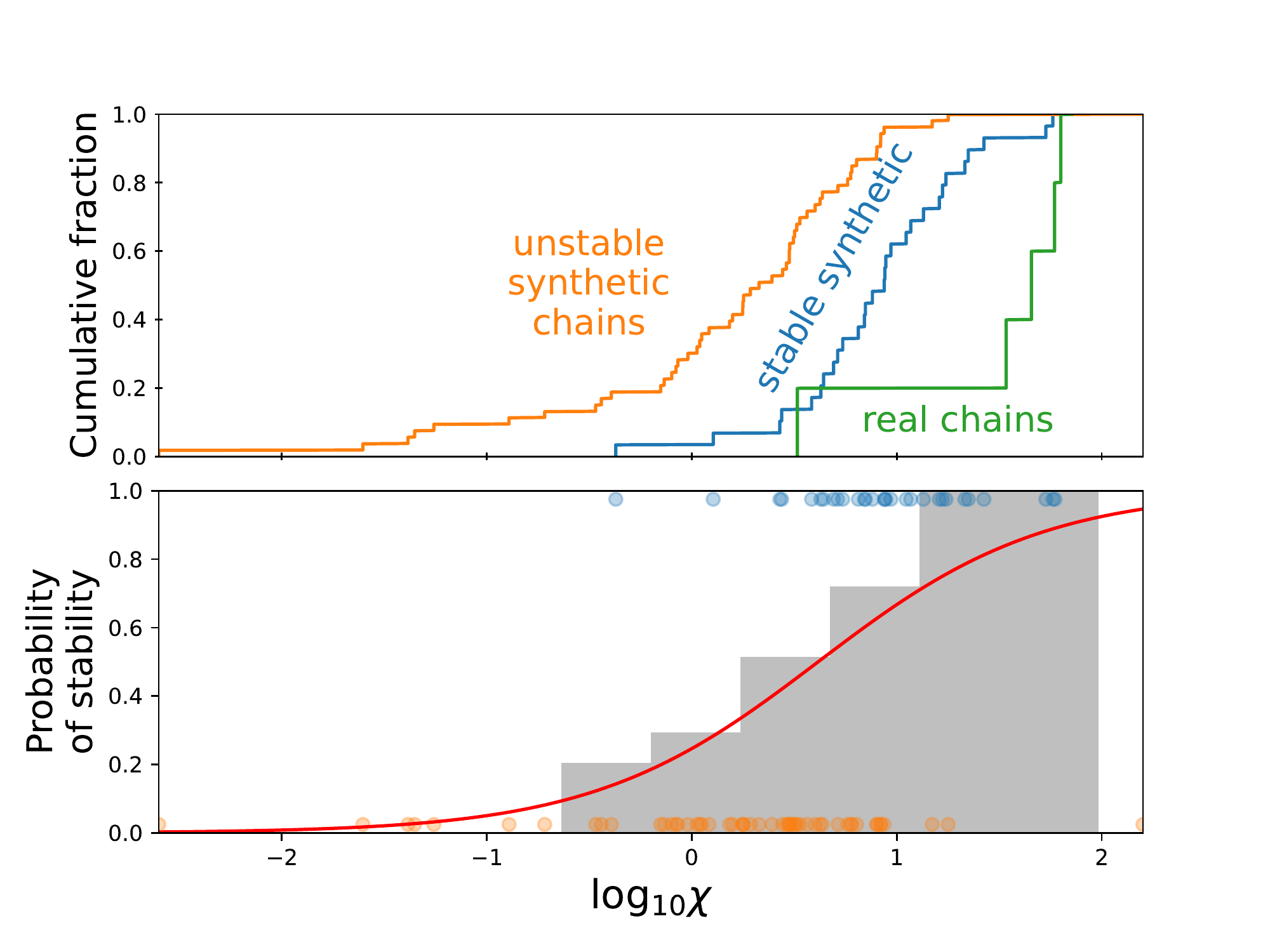}
    \caption{Relationship between our stability criterion $\chi$ and the true stability of synthetic and real resonant chains. Top: the cumulative distribution of $\log_{10}\chi$ for unstable and stable synthetic resonant chains from \cite{Izidoro2021}, and the half-dozen well-characterized chains with small planets. Bottom: blue and orange points mark the same synthetic systems as in the top panel. Gray bars show the fraction of systems within that bin that are stable; the red curve is a logistic regression fit of the probability of stability.}
    \label{fig:log_fit}
\end{figure}
Our hypothesis is that the secondary resonance sets a stability threshold of $\chi_\text{crit} \approx 1$, above which the chain is stable. Of the 30 stable systems, 29 have $\chi > 1$. However, the unstable systems have a broader distribution, clustering around $\chi \sim 2$. To quantify the boundary, we used a logistic regression (Figure \ref{fig:log_fit}) to model the probability of stability given only the $\log_{10}\chi$ of the chain. The fit suggests that the threshold is $\chi_\text{crit} \sim 3$. That is, stability over $10^9$ orbits seems to prefer wider spacing between synodic and libration frequencies than our criterion predicts.

Such a result is surprising in the context of our experiments in Section 3, which suggest that $\chi_\text{crit}$ is near unity for $N\lesssim 6$ and \textit{smaller} for higher-multiplicity systems. While fully understanding this discrepancy is outside the scope of our work, we can speculate on possible sources. Our resonant chains formed in Section 2 are especially ``clean,'' that is, all two-body and three-body resonant angles librate with small amplitudes. In contrast, larger libration amplitudes in the \cite{Izidoro2021} systems could render them more vulnerable to higher-order secondary resonances that appear at $\chi>1$. Another possible explanation is that modulation of the resonant frequencies and widths by secular interactions with other planets in the system causes $\chi$ to vary over long timescales \citep{Tamayo2021}.

As for the \textit{observed} set of resonant chains, Figure \ref{fig:log_fit} shows that they generally have $\chi \sim 30 - 100$, with the exception of TRAPPIST-1, for which $\chi \approx 3$. Finally, it is important to note that these simulations only capture the first $\SI{50}{\mega yr}$, but instabilities can occur after billions of orbits \citep{Petit2020}. Some systems recorded as `stable' might actually be unstable with a longer integration time that is representative of the age of typical exoplanet systems.

\section{Discussion}
Inspired by the analytical study of resonant chains, we have identified a criterion to quantify the stability of planets locked in a chain of resonances in accordance with the \cite{Pichierri2020} mechanism. We argue that the overlap between a fast resonant libration frequency and a slow difference of synodic frequencies leads to chaotic behavior and a dynamical instability. Our criterion predicts a maximum planet mass in a chain of $N$ planets with $k$:$k-1$ resonances and closely agrees with numerical simulations for $k>2$ and $N\lesssim 6$.

In addition, this mechanism explains the counterintuitive result found by \cite{Matsumoto2020} wherein an instability in a maximally-packed resonant chain can be triggered by decreasing the planet masses by $10\%$. Specifically, resonant chain formation occurs in a dissipative environment that suppresses the instability. During migration, the maximum libration frequency can approach, or even exceed, the slowest synodic frequencies, but the system settles into a local island of stability with $\chi < 1$. After the disk is removed and the masses are decreased, the libration frequencies change and the system enters the chaotic region between the island of stability and the $\chi=1$ boundary. While initially discovered numerically, mass loss may in fact be a plausible candidate for the trigger of dynamical instabilities in packed resonant chains. Indeed, mass loss of this magnitude is physically reasonable as a result of photoevaporation \citep{Owen2019}. Furthermore, the highest libration frequency typically comes from the innermost resonance and hence depends only on the masses of the inner two planets. Those planets are most susceptible to photoevaporation by virtue of their proximity to the star.

Machine learning models have been especially successful in analyzing the stability of multi-planet systems. To compare our results to previous work, we used the state-of-the-art SPOCK model \citep{Tamayo2020} to predict the stability of the synthetic chains presented in Section 4. We set the probability threshold to be 0.5 and ran the model in two different ways. First, we use as input each system in its entirety from \cite{Izidoro2021}, only excluding the planets below $0.1M_\oplus$.
SPOCK correctly predicted the stability of 13 of the 30 stable systems and 42 of the 54 unstable ones. Second, we input each of the individual resonant chains identified in the systems of \cite{Izidoro2021}, as described in Section 4, and use SPOCK to compute the probability of their stability. We treat those as independent random variates, and for each system we estimated the probability of stability of the whole system by computing the probability that every chain within it is stable. In that case, SPOCK correctly predicted the stability of 20 of the 30 stable systems and 31 of the 54 unstable ones. As a comparison, our one-dimensional logistic regression (Figure \ref{fig:log_fit}) achieves $25/30$ for stable systems and $40/54$ for unstable with the same probability threshold. Accordingly, the specific problem of resonant chain stability constitutes an instance where a careful analytical treatment is comparable to or surpasses general supervised machine learning techniques.

If it is indeed true as some have suggested that non-resonant systems of small planets are the products of instabilities, the mechanism of instability is of considerable importance. Previous suggestions include changes in the stellar $J_2$ moment \citep{Spalding2016}, stellar or planetary mass loss \citep{Matsumoto2020}, or a simple overpacking of the system during the disk phase \citep{Izidoro2017}. Our work does not rule out any of these mechanisms, but clarifies the dynamics underpinning the onset of the instability. Future work should explore the consequences of each of these instability mechanisms to determine whether they leave signatures detectable in the planet population.

\section*{Acknowledgements}
We are grateful to Antoine Petit and an anonymous referee for valuable feedback that significantly improved this work. We thank Gabriele Pichierri and Sean Raymond for insightful discussions and Andr\'{e} Izidoro for providing simulation results. K. B. is grateful to Caltech, the Caltech Center for Comparative Planetary Evolution, the David and Lucile Packard Foundation, and the Alfred P. Sloan Foundation for their generous support. A. M. acknowledges support from the ERC advanced grant HolyEarth N. 101019380.

\bibliography{main}
\bibliographystyle{cas-model2-names}

\end{document}